\documentclass[aps,pre,twocolumn,superscriptaddress,nobibnotes,nodoi,noeprint]{revtex4-2}
\usepackage{amsmath,amssymb,physics,bm,siunitx}
\usepackage{xcolor,graphicx}
\usepackage{soul}
\usepackage[colorlinks=true,citecolor=blue,urlcolor=blue,linkcolor=black]{hyperref}
\usepackage{setspace}

\newcommand{\rms}{\rm\scriptscriptstyle}
\newcommand{\rhoL}{\rho_{\rm\scriptscriptstyle L}}
\newcommand{\rhoR}{\rho_{\rm\scriptscriptstyle R}}
\newcommand{\rhoB}{\rho_{\rm\scriptscriptstyle B}}
\newcommand{\jB}{j_{\rm\scriptscriptstyle B}}
\newcommand{\rhoonemax}{\rho_1^{\rm max}}
\newcommand{\rhotwomax}{\rho_2^{\rm max}}
\newcommand{\wsat}{w_{\rm sat}}
\DeclareMathOperator*{\argmin}{argmin}
\DeclareMathOperator*{\argmax}{argmax}
\newcommand{\GammaL}{\Gamma_{\rm L}}
\newcommand{\GammaR}{\Gamma_{\rm R}}

\begin{document}

\title{Weak pinning and long-range anticorrelated motion of phase boundaries\\ in driven diffusive systems}

\author{S\"{o}ren~Schweers}
\email{sschweers@uos.de}
\affiliation{Universit\"{a}t Osnabr\"{u}ck, Fachbereich Mathematik/Informatik/Physik, Barbarastra{\ss}e 7, D-49076 Osnabr\"uck, Germany}

\author{David F. Locher}
\email{david.locher@rwth-aachen.de}
\affiliation{Universit\"{a}t Osnabr\"{u}ck, Fachbereich Mathematik/Informatik/Physik, Barbarastra{\ss}e 7, D-49076 Osnabr\"uck, Germany}
\affiliation{Institute for Quantum Information, RWTH Aachen University, 52056 Aachen, Germany}

\author{Gunter M. Schütz} 
\email{gschuetz04@yahoo.com}
\affiliation{Departamento de Matem\'{a}tica, Instituto Superior T\'{e}cnico, Universidade de Lisboa, Av. Rovisco Pais 1, 1049-001 Lisbon, Portugal}

\author{Philipp Maass} 
\email{maass@uos.de}
\affiliation{Universit\"{a}t Osnabr\"{u}ck, Fachbereich Mathematik/Informatik/Physik, Barbarastra{\ss}e 7, D-49076 Osnabr\"uck, Germany}

\date{\today}

\begin{abstract}
We show that domain walls
separating coexisting extremal current phases in driven diffusive systems exhibit complex stochastic dynamics, with a subdiffusive temporal growth of position fluctuations due to long-range anticorrelated current fluctuations and a weak pinning at long times. This weak pinning manifests itself in a saturated width of the domain wall position fluctuations that increases sublinearly with the system size. As a function of time $t$ and system size $L$, the width $w(t,L)$ exhibits a scaling behavior $w(t,L)=L^{3/4}f(t/L^{9/4})$, with $f(u)$
constant for $u\gg1$ and $f(u)\sim u^{1/3}$ for $u\ll1$.  An Orstein-Uhlenbeck process with long-range anticorrelated noise is shown to capture this scaling behavior. Results for the drift coefficient of the domain wall motion point to memory effects in its dynamics.
\end{abstract}

\maketitle

Driven diffusive systems of interacting particles exhibit a large variety of nonequilibrium structures
\cite{Spohn:1991, Schmittmann/Zia:1995}. Fundamental aspects of them can be understood from the study
of driven lattice gas (DLG) models 
\cite{Schmittmann/Zia:1995, Derrida:1998, Schuetz:2001, Frey/etal:2004, Blythe/Evans:2007, Schadschneider/etal:2010, Kriecherbauer/Krug:2010, Chou/etal:2011, Popkov/etal:2015, Mallick:2015, Fang/etal:2019}.
These models have found numerous applications, ranging from microscale processes, such as ionic, colloidal, and 
molecular transport through narrow channels \cite{Kukla/etal:1996, Wei/etal:2000, Hille:2001},
ribosome translation along mRNA strands \cite{MacDonald/etal:1968, Ciandrini/etal:2010, Klumpp/Hwa:2008, Zia/etal:2011, Erdmann-Pham:2020, Keisers/Krug:2023}, cargo transport and
motility of molecular motors \cite{Lipowsky/etal:2010, Chowdhury:2013, Kolomeisky:2015, Jindal/etal:2020}, 
or interface  growth \cite{Krug:1997, Takeuchi/Sano:2010, HalpinHealey/Takeuchi:2015, Spohn:2017, Quastel/Sarkar:2023} to macroscopic processes subject to randomness
like vehicular traffic \cite{Nagel/Schreckenberg:1992, Chowdhury/etal:2000, Helbing:2001, Maerivoet/DeMoor:2005, Schadschneider/etal:2010, Gupta/etal:2023}.

In DLGs with boundaries open to particle reservoirs, phase transitions between nonequilibrium steady states occur even 
in one-dimension \cite{Krug:1991, Kafri/etal:2003}. 
They manifest themselves in 
a singular behavior of the bulk particle density $\rhoB$
as a function of the control parameters, which are
the rates of particle injection from and ejection into the reservoirs. By applying bulk-adapted couplings 
\cite{Antal/Schuetz:2000, Dierl/etal:2012, Dierl/etal:2013}, all possible nonequilibrium phases 
can be inferred from the principle 
of extremal current \cite{Popkov/Schuetz:1999}:
for particle densities $\rhoL$ and $\rhoR$ of 
reservoirs at the left and right boundary, the bulk density $\rhoB$ in the system's interior is
\begin{equation}
\rhoB=\left\{\begin{array}{l@{\hspace{1em}}l}
\displaystyle\argmin_{\rhoL\le\rho\le\rhoR}\{\jB(\rho)\}\,,
& \rhoL\le\rhoR\,,\\[3ex]
\displaystyle\argmax_{\rhoR\le\rho\le\rhoL}\{\jB(\rho)\}\,,
& \rhoR\le\rhoL\,,
\end{array}\right.
\label{eq:extremal-current-principles}
\end{equation}
where
$\jB(\rho)$ is the current in the nonequilibrium steady state (NESS) of a closed system, 
as, e.g., obtained for periodic boundary conditions.
Equation~\eqref{eq:extremal-current-principles} implies that
if $\jB(\rho)$ has a local extremum at some $\rho_{\rm ext}$, an extremal current phase
with $\rhoB=\rho_{\rm ext}$ must occur.
These phases are particularly interesting because they are determined by the 
intrinsic dynamics:
the bulk density is controlled but not 
induced by the coupling to reservoirs.

The dynamics of phase boundaries between coexisting nonequilibrium phases with different $\rhoB$ 
have been studied in the prototype DLG \cite{Kolomeisky/etal:1998, Parmeggiani/etal:2003, Roy/etal:2020}, the 
asymmetric simple exclusion process (ASEP) and variants of it, where particle interactions are
solely due to site exclusion, i.e.\ a lattice site can be occupied by at most one particle .
In the standard ASEP, two boundary-induced phases 
with $\rhoB=\rhoL$ and
$\rhoB=\rhoR$ coexist at $\rhoL=1-\rhoR\le1/2$ \cite{Schuetz/Domany:1993, Derrida/etal:1993}
and the domain wall (DW) separating them performs a random walk with reflecting boundaries 
\cite{Kolomeisky/etal:1998}.  

Here we unravel the dynamical behavior of phase boundaries between coexisting extremal current phases.
A coexistence of two maximal (or minimal) current phases requires the existence of two local maxima (minima) at densities $\rho_1\neq\rho_2$ with $\jB(\rho_1)=\jB(\rho_2)$. A representative model having a 
current-density relation with two degenerate local maxima is the ASEP with repulsive nearest neighbor-interactions between  particles \cite{Katz/etal:1984, Popkov/Schuetz:1999, Dierl/etal:2012, Dierl/etal:2013}.

We show in this Letter that the DW separating the
two maximal current phases exhibits an intriguing 
dynamical and localization behavior: the fluctuations of its position have a width $w$ increasing
subdiffusively with time due to long-term anticorrelations of local current fluctuations, $w\sim t^{1/3}$.
At long times, $w$ saturates at a value $\wsat$ increasing sublinearly with the system size $L$, $\wsat\sim L^{3/4}$.
This implies that the DW in the saturated regime
appears randomly over an infinite region in the thermodynamic limit, while it covers a zero fraction of the system,
as the relative saturated width
$\wsat/L\sim L^{-1/4}$ goes to zero for $L\to\infty$. We refer to this localization behavior as weak pinning of the DW. 
As a function of $t$ and $L$, $w(t,L)$, obeys the scaling law 
\begin{equation}
w(t,L)=L^{\alpha}f\left(t/L^{z}\right)\,,
\label{eq:w-scaling}
\end{equation}
where $\alpha=3/4$, $z=9/4$, and
$f(u)\sim u^\beta$ with $\beta=1/3$ for $u\ll1$, and $f(u)\sim {\rm const.}$ for $u\gg1$.
A Langevin equation is set up, which describes the scaling behavior of $w(t,L)$.

\begin{figure}[t!]
\centering
\includegraphics[width=\columnwidth]{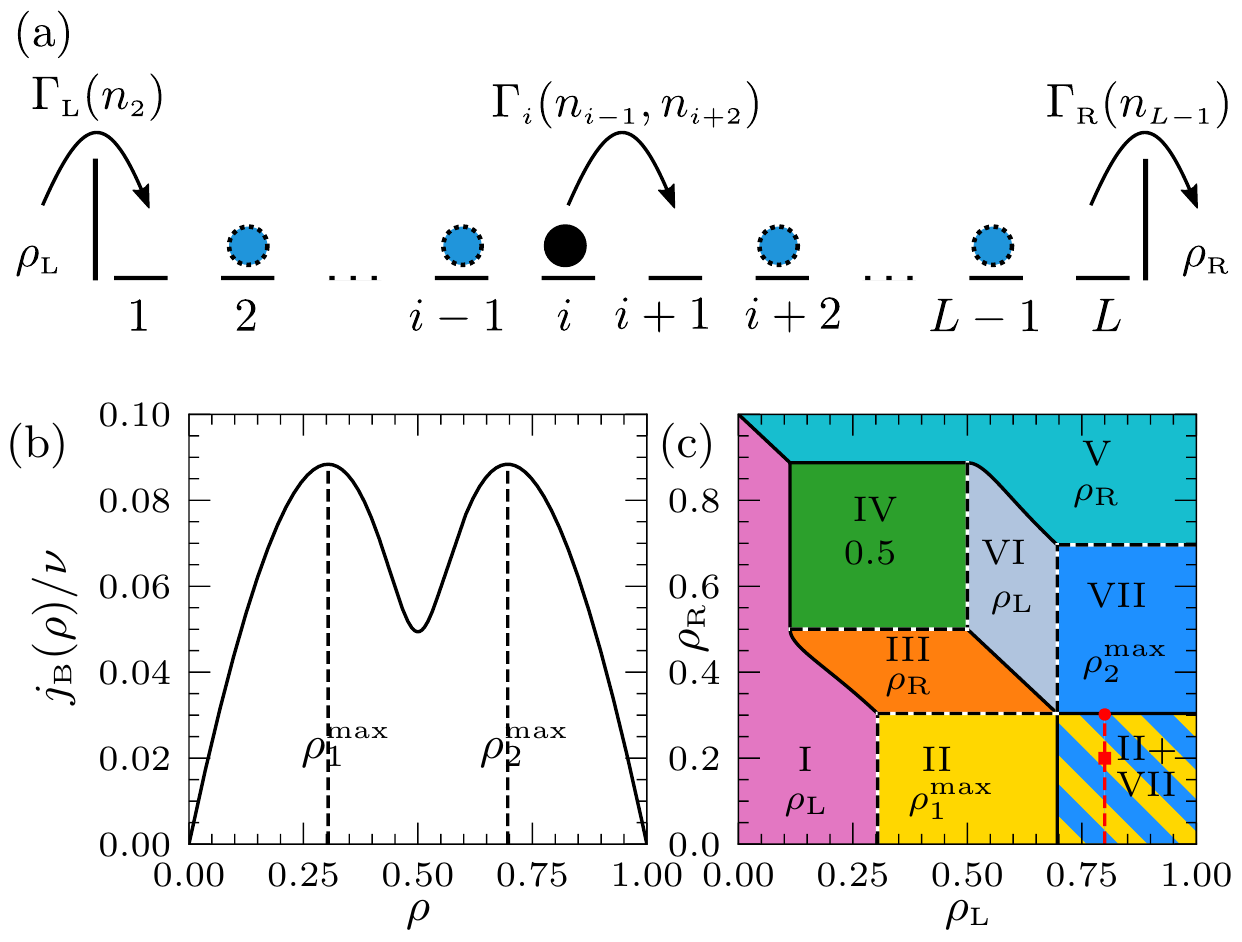}
\caption{(a) Illustration of the open TASEP with repulsive nearest-neighbor interaction $V$: Particles are injected with
rate $\Gamma_{\rms L}(n_2)$ from the left particle reservoir, ejected with rate $\Gamma_{\rms R}(n_{L-1})$ 
to the right reservoir, and
performing unidirectional jumps with rates $\Gamma_i(n_{i-1},n_{i+2})$ \eqref{eq:Glauberrates}
inside the system. (b) Current-density relation for $V=2V_c=4\ln(3)$ in the closed TASEP.
(c) Nonequilibrium phase diagram
following from (b) by applying the extremal current principle~\eqref{eq:extremal-current-principles}.
In the  yellow-blue shaded square the two maximum current phases II and VII coexist. The red square, circle and line
in this area mark reservoir densities, for which DW dynamics were simulated.}
\label{fig:model_and_bulk_properties}
\end{figure}

Figure~\ref{fig:model_and_bulk_properties}(a) illustrates the model, where for simplicity we 
consider the totally asymmetric simple exclusion process (TASEP) with nearest-neighbor interactions 
\footnote{This is not a relevant simplification as discussed in Ref.~\cite{Dierl/etal:2013}.} and open boundaries.
The particles perform unidirectional jumps between neighboring lattice sites $i$, $i=1,\ldots,L$, where the jump rate
$\Gamma_i(n_{i-1}, n_{i+2})$ from a site $i$ to a vacant site $(i+1)$ is given by the Glauber rate
\begin{equation}
\Gamma_i(n_{i-1}, n_{i+2}) = \frac{\nu}{\exp[V(n_{i+2}-n_{i-1})]+1}\,.
\label{eq:Glauberrates}
\end{equation}
Here, $\nu$ is an attempt frequency and $V$ the repulsive nearest-neighbor interaction in units of the thermal energy;
$n_i$ are occupation numbers, i.e.\ $n_i=1$ if site $i$ is occupied and $n_i=0$ otherwise. 
We use $\nu^{-1}$ as the time unit, $\nu=1$. The length unit is given by the lattice constant.
The particle injection and ejection rates $\Gamma_{\rms L}(n_2)$ and $\Gamma_{\rms R}(n_{L-1})$
at the left and right boundaries are bulk-adapted \cite{Hager/etal:2001, Dierl/etal:2013}. 
Further details of the model are given
in Supplemental Material (SM), see below.

\begin{figure}[t!]
\centering
\includegraphics[width=\columnwidth]{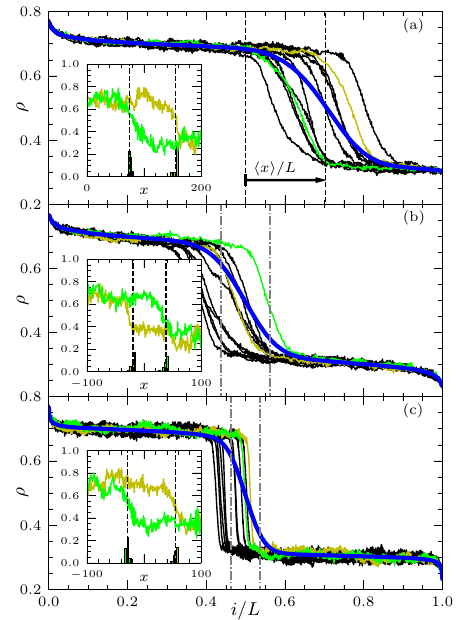}
\caption{Simulated profiles of the particle density in the coexistence region II+VII 
at randomly selected times (thin solid lines) for
(a) $(\rhoL,\rhoR)=(0.8,0.3016)$ and $L=500$, (b) $(\rhoL,\rhoR)=(0.8,0.2)$ and $L=500$, and 
(c) $(\rhoL,\rhoR)=(0.8,0.2)$ and $L=4000$. The thick blue lines mark the long-time averaged 
densities. The dashed vertical lines in (a) indicate the displacement $\langle x\rangle$ 
of the mean DW position from the center. The dash-dotted vertical lines in (b) and (c)
indicate the standard deviation $\pm w_{\rm sat}/L$ of the saturated DW position fluctuations around the
center. The insets show two short-time averages of the fluctuating particle density (noisy green and olive curves), 
as well as histograms of the distribution of ghost particle positions. Their mean value
marks the microscopic position of the DW (dashed vertical lines; $x=i-L/2$ is the position with respect to the center).}
\label{fig:DensityProfile}
\end{figure}

Above a critical value $V_c=2\ln3$, the current density relation $\jB(\rho)$ in a closed system has a double hump structure, where
$\rho=\langle n_i\rangle$ with $\langle\ldots\rangle$ the average in the NESS.
The current density relation is known exactly \cite{Hager/etal:2001, Dierl/etal:2013} and shown
 in Fig.~\ref{fig:model_and_bulk_properties}(b) for $V=2V_c$.
It has maxima at $\rhoonemax=(1-[3-\sqrt{8e^V/(e^V-1)}]^{1/2})/2\cong0.304$
and $\rhotwomax=1-\rhoonemax\cong0.696$. 
Applying the extremal current principle~\eqref{eq:extremal-current-principles} yields
the phase diagram in Fig.~\ref{fig:model_and_bulk_properties}(c). In the whole yellow-blue shaded square
$\{(\rhoL,\rhoR)\,|\, \rhotwomax\!<\!\rhoL\!<\!1, \; 0\!<\!\rhoR\!<\!\rhoonemax\}$ 
at the lower right corner of the phase diagram,
the two maximal current phases~II and VII with $\rho=\rhoonemax$ and $\rhotwomax$ coexist in the open system.
The phases become visible on a microscopic level by looking at many particle trajectories left and right of the DW. An example 
is given in SM.

Figures~\ref{fig:DensityProfile}(a)-(c) show representative density profiles of the coexisting phases as a function
of the normalized position $i/L$
at various times for different $L$ and different pairs $(\rhoL,\rhoR)$ of reservoir densities in the coexistence region II+VII. These profiles have been determined from kinetic Monte Carlo (KMC) simulations by averaging the occupation numbers in a time window $\Delta t=5\times10^4$ at various times. They decay from $\rhotwomax$ to $\rhoonemax$. 
At different times in the interval $\Delta t$, the DW occurs at different random positions, leading to a
decay of the profile over a region larger than the intrinsic width $\xi$ of the DW.
Profiles averaged over long times are represented by the thick blue lines in Figs.~\ref{fig:DensityProfile}(a)-(c) and show the stationary mean position of decay.
Videos of the time-dependent density profiles are provided in SM.

In Figs.~\ref{fig:DensityProfile}(a) and \ref{fig:DensityProfile}(b),
the system size $L=500$ is the same, but the reservoir densities are different, with (a) $(\rhoL,\rhoR)=(0.8,0.3016)$ very 
close to the boundary to phase VII [see Fig.~\ref{fig:model_and_bulk_properties}(c)], and 
(b) $(\rhoL,\rhoR)=(0.8,0.2)$ on the ``symmetry line'' $\rhoR=1-\rhoL$.
While the mean position of the DW is in the system's center at $i/L\cong1/2$ in (b), it is shifted to the right in (a).
In Fig.~\ref{fig:DensityProfile}(c) the reservoir densities are the same as in Fig.~\ref{fig:DensityProfile}(b), but the system 
size is eight times larger, $L=4\times 10^3$. For the larger system size in (c), 
the position fluctuations of the DW relative to the system size become smaller.

\begin{figure}[t!]
\centering
\includegraphics[width=\columnwidth]{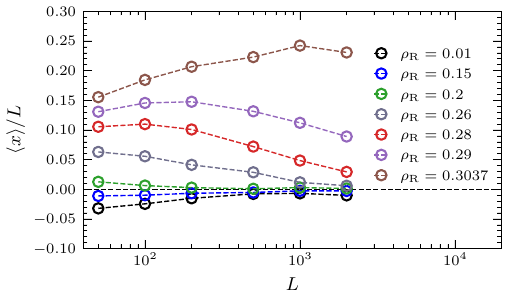}
\caption{Mean DW position $\langle x\rangle$ as a function of $L$ for
fixed $\rhoL=0.8$ and various $\rhoR$ corresponding to points on the red line shown in Fig.~\ref{fig:model_and_bulk_properties}(c).}
\label{fig:DWposition}
\end{figure}

To quantify these effects, we have determined the instantaneous DW position by
adding $N_{\rm g}=40$ ghost particles to the system, which neither interact with themselves nor affect the stochastic dynamics of the
regular particles.
After each jump of a regular particle, 50\% of the ghost particles are randomly selected. 
If a selected ghost particle is on a site occupied (not occupied)
by a regular particle it is moved one site to the right (left). Due to these dynamics, the ghost particles accumulate at the DW position
after a transient time. Thereafter the DW position $x$ is accurately given by the mean position
\begin{equation}
x(t)=\frac{1}{N_g}\sum_{i=1}^{N_g}x_i(t)
\end{equation}
of all ghost particles. We specify this position with respect to the system's center, i.e.\ the position $x_i$ of ghost particle $i$ in the sum 
is $x_i=j-L/2$ if it occupies site $j$.

Histograms of ghost particle positions are given in the insets of Figs. 2(a)-(c), together with density profiles of the regular particles. The stronger spatial fluctuations of these profiles compared to those in the main figure are due to averaging of occupation numbers over a shorter time window $10^{-2} \Delta t$, in which the DW movement can be neglected.
The standard deviation of the ghost particle positions is a 
measure of the intrinsic width $\xi$ of the DW, or, differently speaking, of the unavoidable uncertainty 
in defining its instantaneous 
position (as some time-averaging is always necessary to obtain sufficiently smooth density profiles). 
We find that $\xi$ is about four 
lattice constants independent of the reservoir densities and $L$, thus demonstrating that the DW is microscopically 
sharp. This is reminiscent of the microscopic sharpness of a shock discontinuity in the open ASEP 
\cite{Krebs/etal:2003,Parmeggiani/etal:2003,Schuetz:2023}. In videos provided in the SM, we show the stochastic change of density profiles in time together with the coupled motion of the ghost particle cloud.
 
\begin{figure}[t!]
\includegraphics[width=\columnwidth]{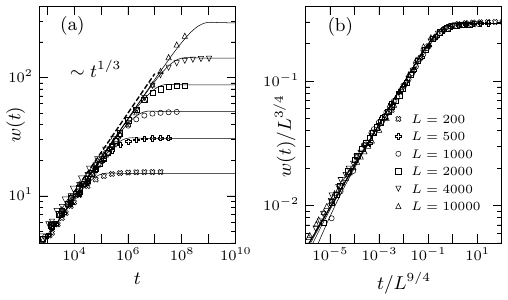}
\caption{(a) Width $w(t,L)$ of DW position fluctuations [see Eq.~\eqref{eq:wdef}] as a function of time
for various system sizes $L$.
(b) Scaled width $w(t,L)/L^{3/4}$ as a function of scaled time $t/L^{9/4}$ demonstrating the scaling 
law \eqref{eq:w-scaling}. The legend in (b) applies to (a) also. Solid lines indicate the results 
from the Ornstein-Uhlenbeck process given by Eq.~\eqref{eq:langevin-model1} with parameters
$\kappa_0=1.133$ and $C_\infty=0.0225$.}
\label{fig:DW_width_and_scaling}
\end{figure}

Figure~\ref{fig:DWposition} shows the stationary mean DW position $\langle x\rangle$
as a function of the system size $L$ for reservoir densities
$(\rhoL=0.8,\rhoR)$ along the red line indicated in the coexistence region II+VII of the phase diagram in Fig.~\ref{fig:model_and_bulk_properties}(c).
For small $L$, $\langle x\rangle$ is displaced from the center of the system, except for $\rhoR=1-\rhoL=0.8$. 
In particular, when $\rhoR$ approaches
$\rhoonemax$, the displacement increases at fixed $L$. This can be explained by the fact that when $\rhoR$
crosses $\rhoonemax$ (at fixed $\rhoL=0.8$), the NESS becomes the maximal current phase VII with 
$\rhoB=\rhotwomax$. Thus
phase~II is increasingly expelled from the system as $\rhoR$ approaches $\rhoonemax$. However, when increasing $L$ at fixed $(\rhoL,\rhoR)$,
the displacement of the mean DW position from the center decreases and approaches zero for large $L$. 
This means that the displacement indicated in Fig.~\ref{fig:DensityProfile}(a)
is a finite-size effect.

We now analyze the stationary fluctuations of the DW position for $(\rhoL,\rhoR)=(0.8,0.2)$, 
where $\langle x\rangle=0$ for all $L$. To this end 
we define the
width of these fluctuations in a system of size $L$ by the standard deviation of the displacement of the instantaneous DW position $x(t)$ from the mean position $\bar{x}(t)=\sum_{t^\prime=0}^{t}x(t^\prime)/(t+1)$ after time $t$,
\begin{equation}
w(t,L)=\sqrt{\frac{1}{(t+1)}\sum_{t^\prime=0}^{t}[x(t^\prime)-\bar{x}(t^\prime)]^2}\,.
\label{eq:wdef}
\end{equation}

Figure~\ref{fig:DW_width_and_scaling}(a) shows $w(t,L)$ 
as a function of time for various $L$.
At small $t$, $w(t,L)\sim t^{1/3}$ is independent of $L$, while for large times, $w$ saturates at a value $w_{\rm sat}\sim L^{3/4}$. When plotting the data in scaled form according to the scaling law \eqref{eq:w-scaling}, they collapse onto one master curve, see Fig.~\ref{fig:DW_width_and_scaling}(b).

How can we explain the scaling behavior of $w(t,L)$? To answer this question, 
we consider the velocity of shock fronts in kinematic 
wave theory \cite{Lighthill/Whitman:1955}, 
which, as a consequence of particle number conservation, is 
$v=[j(\rho_2)-j(\rho_1)]/(\rho_2-\rho_1)$, where $\rho_2$ ($\rho_1$) are the particle densities before (after) the shock front.
Applying this law to particle densities $\rho_\pm(x)\simeq \rho(x\pm\xi)$ 
and fluctuating currents $\jB(\rho_\pm(x))+\delta j_\pm(t)$  right and left of the DW, we can write
\begin{equation}
\frac{\dd x}{\dd t} = \frac{\jB(\rho_-(x))\!-\!\jB(\rho_+(x))\!+\!\delta j(t)}{\rho_-(x)-\rho_+(x)}
\simeq \frac{\delta j(t)}{\rhotwomax\!-\!\rhoonemax}\,.
\label{eq:interfaceV}
\end{equation} 
Here we have set $\rho_-(x)\simeq\rhotwomax$, $\rho_+(x)\simeq\rhoonemax$, and
$\delta j(t)=\delta j_-(t)-\delta j_+(t)$. Current fluctuations in NESS of nonreversible DLG
 are anticorrelated: their 
correlation function $C(t)=\langle \delta j(t)\delta j(0)\rangle$ decays as $C(t)\sim -t^{-4/3}$,
and the integral over $C(t)$ is zero \cite{Ferrari/Spohn:2016}. This leads to the subdiffusive $t^{1/3}$-scaling
of $w(t,L)$, see SM.

As for the weak pinning of the DW, we argue that it is caused
by the tendency of the density profiles left and right from the DW
to match profiles of the corresponding maximal current phases. 
The functional form of these matching profiles is universal
and decays 
as a power law $\sim 1/\sqrt{l}$ with the distance $l$ from the boundary \cite{Krug:1991,Hager/etal:2001}. 
Thus, close to $x=0$ the
mean densities left and right of the DW become slightly different from $\rhotwomax$ and $\rhoonemax$, giving rise 
to a symmetric restoring force of the DW position towards $x=0$. As gradients of the mean densities
are very small and decrease with $L$, a linear approximation $\delta\rho(x)=\delta\rho'(0)x$ should be appropriate. This
suggests that the restoring force is linear with a strength decreasing with $L$. In view of the 
characteristic time scale $\sim L^{9/4}$ found in the simulations, we thus write
\begin{equation}
\frac{\dd x}{\dd t}=-\frac{\kappa_0}{L^{9/4}}\,x+\eta(t)\,.
\label{eq:langevin-model1}
\end{equation}
Here $\kappa_0>0$ is a constant and $\eta(t)$ a stationary Gaussian process with zero mean and
autocorrelation function $C(t)$ with $\int_0^\infty\dd t\, C(t)=0$ and asymptotic behavior 
$C(t)\sim -C_\infty (rt)^{-4/3}$ for $t\to\infty$, where $C_\infty>0$ is a constant 
and $r^{-1}$ a microscopic time scale (e.g., the inverse attempt frequency $\nu^{-1}=1$ in the TASEP model).  

Equation~\eqref{eq:langevin-model1} describes an Ornstein-Uhlenbeck process with long-term anticorrelated noise.
As shown in SM, this yields a scaling behavior of $w(t,L)$ in agreement with
Eq.~\eqref{eq:w-scaling}, where
the scaling function $f(u)$ behaves as $f(u)\sim 3C_\infty^{1/2} r^{-2/3} u^{1/3}$ for $u\ll1$ and 
$f(u)\sim {\rm const.}=[3\Gamma(\tfrac{2}{3})C_\infty]^{1/2}r^{-2/3}\kappa_0^{-1/3}$for $u\gg 1$ [$\Gamma(.)$ is the Gamma 
function]. The simulated data can be well fitted by this 
model, see the solid lines in Fig.~\ref{fig:DW_width_and_scaling}.

We have also verified the presence of a linear restoring force by determining the drift coefficient 
$D_1=\langle [x(t+\tau)-x(t)]\,|\,x(t)=x\rangle/\tau$ from the simulations, see SM. 
As the DW is well defined only on a coarse-grained time scale, we cannot take the limit $\tau\to0$ and
have analyzed $D_1$ for different $\tau$ \cite{Anteneodo/etal:2010}. For the Ornstein-Uhlenbeck process in 
Eq.~\eqref{eq:langevin-model1}, $D_1=D_1(x,L;\tau)=-x(1-e^{-\kappa\tau})/\tau$ with
$\kappa=\kappa_0/L^{9/4}$. Our simulation results are in agreement with this prediction,
but with a $\kappa$ depending on $\tau$, approaching 
a constant $\propto L^{-9/4}$ for large $\tau$. 
This means that Eq.~\eqref{eq:langevin-model1} does not provide
a complete description of the DW dynamics. We believe that the shape restoring of the density profiles 
is affected by memory effects not included in Eq.~\eqref{eq:langevin-model1}.

In summary, we have shown for the TASEP with repulsive nearest-neighbor interaction that the DW separating extremal current phases is microscopically sharp and exhibits surprisingly rich stochastic dynamics. A subdiffusive growth of the variance of the DW position arises from anticorrelated current fluctuations. At large times $t\gg L^{9/4}$ the DW becomes weakly pinned in a region increasing sublinearly 
with the system size. The crossover exponent $9/4$ defines a new dynamical exponent for relaxation processes in DLGs. We interpret this weak pinning by the tendency of the density profiles left and right of the DW to
match their preferred shapes. On macroscopic scale, i.e., by rescaling the lattice by $1/L$, the 
density profile has two constant segments of densities $\rhotwomax$ and $\rhoonemax$ respectively. The DW marking the transition point from $\rhotwomax$ to $\rhoonemax$ 
becomes sharp and corresponds to a so-called contact discontinuity \cite{Bressan:1999}. 
In contrast to the microscopically well-understood shock discontinuities
appearing in the ASEP on the coexistence line, see e.g.
\cite{Ferrari/etal:1991,Krebs/etal:2003,Derrida/etal:1997,Balazs/etal:2010,Belitsky/Schuetz:2013}, 
little is known about the microscopic structure of contact discontinuities. 
Our findings constitute a first step towards their systematic exploration.

Because these phenomena occur on large time and length scales, 
they are independent of microscopic details and therefore expected to
hold true in general for driven diffusive systems with short-range particle interactions. 
They imply that DW dynamics between extremal current phases reflect current correlations and slow power-law decays of density profiles in extremal current phases. Investigations of DW dynamics can thus be used as a dynamical probe for analyzing these pertinent features of driven diffusive systems.

\begin{acknowledgments}
This work has been funded by the Deutsche Forschungsgemeinschaft (DFG, Project No.\ 355031190). We sincerely thank A.\ Zahra, V.\  Popkov, and the members of the DFG Research Unit FOR2692 for fruitful discussions.
\end{acknowledgments}


%

\onecolumngrid
\newpage
\renewcommand{\theequation}{S\arabic{equation}}
\renewcommand{\thefigure}{S\arabic{figure}}
\setcounter{equation}{0}
\setcounter{figure}{0}

\begin{center}
\setcounter{page}{1}
{\large\bf Supplemental Material for}\\[2ex]
{\large\bf Weak pinning and long-range anticorrelated motion of phase boundaries\\ in driven diffusive systems}\\[2ex]
S\"{o}ren~Schweers,$^1$ David F. Locher,$^{1,2}$ Gunter M. Sch\"utz,$^3$ and Philipp Maass$^1$\\[1ex]
$^1$\textit{Universit\"{a}t Osnabr\"{u}ck, Fachbereich Mathematik/Informatik/Physik,\\ Barbarastra{\ss}e 7, D-49076 Osnabr\"uck, Germany}\\[1ex]
$^2$\textit{Institute for Quantum Information, RWTH Aachen University, 52056 Aachen, Germany}\\[1ex]
$^3$\textit{Departamento de Matem\'{a}tica, Instituto Superior T\'{e}cnico, Universidade de Lisboa,\\
Av. Rovisco Pais 1, 1049-001 Lisbon, Portugal}
\end{center}

\setstretch{1.5}
\vspace{1ex}\noindent

In Sec.~\ref{sec:model_and_simulation_details} of this Supplemental Material, we provide details of the driven lattice gas model and the simulation methods, which we used to explore the domain wall motion between extremal current phases in driven diffusive systems. 
Section~\ref{sec:particle-trajectories} presents particle trajectories in the NESS, illustrating
coexisting maximal current phases in the model on a microscopic level. 
The behavior of the domain wall width resulting from the Ornstein-Uhlenbeck process with long-term anticorrelated noise [Eq.~(7) in the main manuscript] is derived in
Sec.~\ref{sec:DW-width-model}. In Sec.~\ref{sec:drift-coefficient} we show simulation
results for the drift coefficient of the domain wall motion. A description of videos of the stochastic
domain wall motion is given in Sec.~\ref{sec:videos}.

\section{Details of the model and kinetic Monte Carlo simulations}
\label{sec:model_and_simulation_details}

The Glauber rates in Eq.~(3) of the main text 
allow for an exact derivation of the current density relation in the bulk \cite{Hager/etal:2001, Dierl/etal:2012, Dierl/etal:2013}.
The result is
\begin{equation}
j_{\rm\scriptscriptstyle B}(\rho)=\nu\left[(\rho-C^{(1)})^2\frac{2f-1}{2\rho(1-\rho)}+(\rho-C^{(1)})(1-f)\right],
\end{equation}
where
\begin{equation}
C^{(1)}=\langle n_i n_{i+1}\rangle_{\rm\scriptscriptstyle eq}
=\frac{1}{2[1\!-\!\exp(-V)]}\left[2\rho\,(1\!-\!\exp(-V))-1+\sqrt{1-4\rho(1\!-\!\rho)[1\!-\!\exp(-V)]}\,\right]
\end{equation}
and 
\begin{equation}
f=\frac{1}{\exp(V)+1}\,.
\end{equation}

\begin{figure}[b!]
\includegraphics[width=\columnwidth]{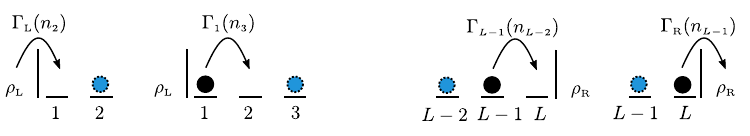}
\caption{Illustration of the injection and ejection rates $\GammaL(n_2)$ and $\GammaR(n_{L-1})$
as well as the rates $\Gamma_1(n_3)$ and $\Gamma_{L-1}(n_{L-2})$ for jumps next to the boundaries.}
\label{fig:illustration_boundary_rates}
\end{figure}

The coupling of the system to reservoirs is bulk-adapted \cite{Hager/etal:2001,Dierl/etal:2012, Dierl/etal:2013}. In this case, the injection rate 
$\GammaL(n_2)$ and ejection rate $\GammaR(n_{L-1}$)
of particles, as well as the rates $\Gamma_1(n_3)$ and $\Gamma_{L-1}(n_{L-2})$ for jumps next 
to the boundaries, see Fig.~\ref{fig:illustration_boundary_rates},
are fully parameterized by the reservoir densities $\rhoL$ and $\rhoR$ (for given interaction $V$):
\begin{subequations}
\label{eq:ba-rates}
\begin{align}
\GammaL(n_2)&= p_{2|2}(01|0n_2;\rhoL,V)\Gamma(0,n_2)+p_{2|2}(11|0n_2;\rhoL,V)\Gamma(1,n_2)\,,
\label{eq:ba-rates-a}\\
\Gamma_1(n_3)&= p_{1|3}(0|10n_3;\rhoL,V)\Gamma(0,n_3)+p_{1|3}(1|10n_3;\rhoL,V)\Gamma(1,n_3)\,.
\label{eq:ba-rates-b}
\end{align}
\end{subequations}
Here, $\Gamma(0,n_2)$, $\Gamma(1,n_2)$ etc.\ are the Glauber rates in Eq.~(3) of the main text. The function
$p_{2|2}(m_1,m_2$ $|m_3\,m_4;\rhoL,V)$ is the probability of finding the occupations
$m_1$ and $m_2$ at sites $i$ and $(i+1)$ for given occupations $(m_3\,m_4)$ at sites $(i+2)$ and $(i+3)$ 
in a bulk system with density $\rhoL$ and interaction $V$.
Likewise, $p_{1|3}(m_1|m_2\,m_3\,m_4;\rhoL,V)$ is the probability of finding an occupation
$m_1$ at site $i$ for given  occupations $(m_2\,m_3\,m_4)$ at sites $(i+1)$, $(i+2)$, and $(i+3)$.
Analogously, the rates $\GammaR(n_{L-1})$ and $\Gamma_{L-1}(n_{L-2})$ are defined.

The bulk-adapted coupling ensures that a semi-infinite system with only one open boundary has a homogeneous 
NESS with constant density given by the boundary densities $\rho_{\rm\scriptscriptstyle L,R}$, which parameterize the 
boundary rates. Kinetic Monte-Carlo simulations of the stochastic jump process
were performed by using the first-reaction algorithm \cite{Gillespie:1978, Holubec/etal:2011}.

\section{Demonstration of the domain wall on a microscopic scale}
\label{sec:particle-trajectories}

\begin{figure}[b!]
\includegraphics[width=0.75\columnwidth]{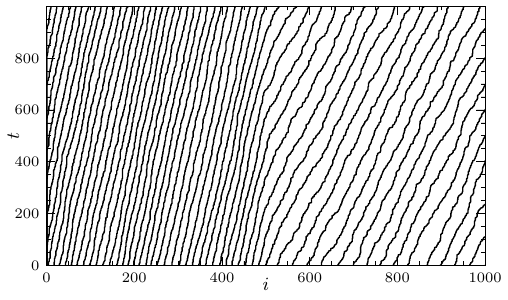}
\caption{Trajectories of every 10th particle are shown for a system of size~$L=10^3$ 
and $\rhoL=0.8$ and $\rhoR=0.2$.}
\label{fig:trajectories}
\end{figure}

The DW in a NESS can be visualized on a microscopic scale by plotting the trajectory of every 10th particle,
see Fig.~\ref{fig:trajectories}. The DW appears close to the center of the system
and remains there during the small observation time 10$^3$ used in this figure. The mean particle distance is much smaller 
to the left of the DW in comparison to the part right of the DW, indicating the coexistence of the two maximal current phases with densities $\rhoonemax$ and $\rhotwomax$. A speedup of particles occurs from a mean velocity $v_<$ to a higher mean velocity $v_>$
when they cross the DW. This effect is a result of particle number conservation, $v_<\rho_<=v_>\rho_>$, i.e.\ with
$\rho_<\cong\rhotwomax$ and $\rho_>\cong\rhoonemax$ it holds $v_>/v_<\cong\rhotwomax/\rhoonemax\cong2.29$. The stationary currents are the same in both domains so that the average speed of the DW vanishes.

\section{Domain wall fluctuations resulting from an Ornstein-Uhlenbeck process with long-term anticorrelated noise}
\label{sec:DW-width-model}
We calculate the width $w(t,L)$ of the distribution of the domain wall position $x(t)$, i.e., its standard deviation,
via a stochastic dynamics given by
\begin{equation}
\frac{\dd x}{\dd t}=-\kappa(L)x+\eta(t) .
\label{eq:langevin1-x}
\end{equation}
Here $\eta(t)$ is a stationary Gaussian process with zero mean and
autocorrelation function $C(t)$ with the properties
\vspace*{-2ex}\begin{subequations}
\begin{gather}
\int\limits_0^\infty\dd t\, C(t)=0\,,\\
C(t)\sim -C_\infty (rt)^{-\gamma}\hspace{0.5em}\mbox{for}\hspace{1em} t\to\infty\,,\hspace{1em}0<\gamma<1\,.
\label{eq:Cproperties}
\end{gather}
\end{subequations}
Here, $r$ is a microscopic rate.
The strength $\kappa(L)$ of the linear restoring force decays with the system size $L$ as
\begin{equation}
\kappa(L)=\kappa_0L^{-\mu}
\label{eq:kappa(L)}
\end{equation}
with an exponent $\mu>0$. For the stochastic DW motion considered in Eq.~(7) of the main text, 
$\gamma=4/3$ and $\mu=9/4$.

For the initial condition $x(0)=x_0$ and a given realization of the noise $\eta(t)$, the solution of Eq.~\eqref{eq:langevin1-x} is
\begin{equation}
x(t)=x_0e^{-\kappa(L)t}+e^{-\kappa(L) t}\int\limits_0^t\dd t' e^{\kappa(L) t'}\eta(t')\,,
\label{eq:x-langevin1-sol}
\end{equation}
from which follows
\begin{subequations}
\begin{gather}
\langle x(t)\rangle=x_0e^{-\kappa t}\,,
\label{eq:xave-langevin1-sol}\\
\langle x^2(t)\rangle=x_0^2e^{-2\kappa t}
+e^{-2\kappa t}\int\limits_0^t\dd t_1\int\limits_0^t\dd t_2\, e^{\kappa (t_1+t_2)}C(|t_1-t_2|)\,.
\label{eq:xsquareave-langevin1-sol}
\end{gather}
\end{subequations}
Equations~\eqref{eq:xave-langevin1-sol} and \eqref{eq:xsquareave-langevin1-sol} yield
\begin{equation}
w^2(t,L)=\langle x^2(t)\rangle-\langle x(t)\rangle^2\label{eq:w2-2}
=e^{-2\kappa t}\int\limits_0^t\dd t_1\int\limits_0^t\dd t_2\, e^{\kappa(t_1+t_2)}C(|t_1-t_2|)\,.
\end{equation}
Substituting the integration variables $t_1$, $t_2$ by $\tau_1=t_1\!-\!t_2$ and $\tau_2=t_1\!+\!t_2$, we obtain
\begin{equation}
w^2(t,L)=e^{-2\kappa t}\int\limits_0^t\dd\tau_1\,C(\tau_1)
\hspace{-0.3em}\int\limits_{\tau_1}^{2t-\tau_1}\hspace{-0.3em}\dd\tau_2\,e^{\kappa\tau_2}
=\frac{2e^{-\kappa t}}{\kappa}\int\limits_0^t\dd t'\,C(t')\sinh[\kappa (t-t')]\,.
\label{eq:w2-gensol}
\end{equation}
For times $1/r\ll t\ll1/\kappa=\kappa_0^{-1}L^\mu$ this behaves as
\begin{align}
w^2(t,L)&\sim 2\int\limits_0^t \dd t'\,C(t')(t-t')\sim -2t\int\limits_t^\infty \dd t'\,C(t')-2\int\limits_0^t \dd t'\,C(t')t'\nonumber\\
&\sim 2C_\infty t\int\limits_0^t \dd t'\, (rt')^{-\gamma}+\frac{2C_\infty}{r}\int\limits^t\dd t'\, (rt')^{2-\gamma}
\sim \frac{2C_\infty}{(\gamma\!-\!1)(2\!-\!\gamma)r^2}\,(rt)^{2-\gamma}\,.
\label{eq:t-scaling}
\end{align}
For times $t\gg 1/\kappa=\kappa_0^{-1}L^\mu$, $w^2(t,L)$ from Eq.~\eqref{eq:w2-gensol}
approaches the constant
\begin{equation}
w^2_{\rm sat}(L)=w^2(t\to\infty,L)=\frac{1}{\kappa}\int\limits_0^\infty \dd t\, C(t)e^{-\kappa t}=\frac{\tilde C(\kappa)}{\kappa}\,,
\label{eq:w2sat}
\end{equation}
where $\tilde C(z)$ is the Laplace transform of $C(t)$. For $z\to0$ this has the asymptotic behavior
\begin{align}
\tilde C(z)&=-\int\limits_0^\infty \dd t\, C(t)(1-e^{-zt})=-\int\limits_0^\infty \frac{\dd u}{z}\, C\!\left(\frac{u}{z}\right)(1-e^{-u})
\sim \frac{C_\infty}{z}\left(\frac{r}{z}\right)^{-\gamma}\int\limits_0^\infty \dd u\, u^{-\gamma}(1-e^{-u})\nonumber\\
&\sim\frac{C_\infty}{(\gamma\!-\!1)r}\left(\frac{z}{r}\right)^{\gamma-1}\left[-u^{-(\gamma-1)}(1-e^{-u})\Biggl|_0^\infty
+\int\limits_0^\infty \dd u\, u^{-(\gamma-1)}e^{-u}\right]
\sim \frac{\Gamma(2\!-\!\gamma)}{\gamma\!-\!1}\frac{C_\infty}{r}\left(\frac{z}{r}\right)^{\gamma-1}\,.
\label{eq:laplaceC-small-z}
\end{align}
Accordingly, Eq.~\eqref{eq:w2sat} yields in the limit of large $L\gg1$, where $\kappa(L)\to 0$:
\begin{equation}
w^2_{\rm sat}(L)\sim 
\frac{1}{\kappa(L)}\frac{\Gamma(2\!-\!\gamma)}{\gamma\!-\!1}\frac{C_\infty}{r}\left(\frac{\kappa(L)}{r}\right)^{-(2-\gamma)}
\sim \frac{\Gamma(2\!-\!\gamma)}{\gamma\!-\!1}\frac{C_\infty}{r^2}\left(\frac{\kappa_0}{r}\right)^{-(2-\gamma)}L^{(2-\gamma)\mu}\,.
\label{eq:L-scaling}
\end{equation}
Equations~\eqref{eq:t-scaling} and \eqref{eq:L-scaling} imply the scaling form
\begin{equation}
w(t,L)=L^{(1-\gamma/2)\mu}\,f\left(\frac{t}{L^\mu}\right)
\label{eq:w-scaling-model1}
\end{equation}
with 
\begin{equation}
f(u)\sim\left\{\begin{array}{cc}
\sqrt{\dfrac{2C_\infty}{(\gamma\!-\!1)(2\!-\!\gamma)r^\gamma}}\,u^{1-\gamma/2}\,, & u\to 0 \hspace{0.5em}(u\ll1)\,,\\[3ex]
\sqrt{\dfrac{\Gamma(2\!-\!\gamma)C_\infty}{(\gamma\!-\!1)\kappa_0^{2-\gamma}r^\gamma}}\,,& u\to\infty \hspace{0.5em}(u\gg1)\,.
\end{array}\right.
\end{equation}
For $\gamma=4/3$ and $\mu=9/4$ this gives
\begin{equation}
w(t,L)=L^{3/4}\,f\left(\frac{t}{L^{9/4}}\right)
\label{eq:w-scaling-model1}
\end{equation}
with 
\begin{equation}
f(u)\sim\left\{\begin{array}{cc}
3\sqrt{C_\infty}r^{-2/3} u^{1/3}\,, & u\to 0 \hspace{0.5em}(u\ll1)\,,\\[1ex]
\sqrt{3\Gamma\!\left(\tfrac{2}{3}\right)C_\infty}\,\kappa_0^{-1/3}r^{-2/3}\,, & u\to\infty \hspace{0.5em}(u\gg1)\,.
\end{array}\right.
\end{equation}

\section{Drift coefficient of stochastic domain wall motion}
\label{sec:drift-coefficient}

We determined the drift coefficient
\begin{equation}
D_1(x,L;\tau)=\langle\left[x(t+\tau)-x(t)\right]|x(t)=x\rangle/\tau
\end{equation}
as a function of $x$ from the KMC simulations for systems of different lengths $L$ and for varying time increment $\tau$. The limit $\tau\to0$ cannot be taken, because some coarse-grained time scale is necessary for the identification of the DW in the discrete lattice model. For the Ornstein-Uhlenbeck process \eqref{eq:langevin1-x}, we obtain from Eq.~\eqref{eq:xave-langevin1-sol} (see also \cite{Anteneodo/etal:2010})
\begin{equation}
D_1(x,L;\tau)=\frac{1}{\tau}\langle [x(t+\tau)-x(t)]\,|\,x(t)=x\rangle=-\frac{x(1-e^{-\kappa\tau})}{\tau}
\label{eq:D1-model1}
\end{equation}
with $\kappa=\kappa_0/L^{9/4}$ from Eq.~\eqref{eq:kappa(L)} ($\mu=9/4$).

\begin{figure}[t!]
\includegraphics[width=0.75\columnwidth]{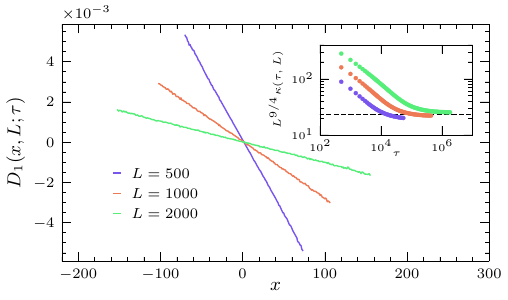}
\caption{Dependence of $D_1(x,L;\tau)$ on $x$ for a fixed $\tau=500$ for different $L$ obtained from the KMC simulations. Data are shown for those $x$, which are taken by the DW sufficiently frequent during the simulated time in the NESS (at least $10^5$ events for obtaining reliable statistical accuracy). 
The inset shows that $L^{9/4}\kappa(\tau,L)$ approaches a constant independent of $\tau$ for large $\tau$.}
\label{fig:drift-coefficient}
\end{figure}

Figure~\ref{fig:drift-coefficient} shows that the linear dependence on $x$ is well confirmed by 
our KMC simulations. The proportionality factor extracted from the slopes $D_1(x,L;\tau)/x$ of the simulated data
is, however, not equal to $-(1-e^{-\kappa\tau})/\tau$ as predicted by \eqref{eq:D1-model1}, but shows a more complex dependence on $\tau$. To describe the behavior, we consider a $\tau$-dependent $\kappa$ 
in Eq.~\eqref{eq:D1-model1}, i.e.\ we define
\begin{equation}
\kappa(\tau,L) = -\frac{1}{\tau}\ln\left(1+\frac{D_1(x,L;\tau)\tau}{x}\right)\,.
\end{equation}
In the inset of Fig.~\ref{fig:drift-coefficient}, we plotted  $L^{9/4}\kappa(\tau,L)$ as a function of $\tau$.
The data indicate that $\kappa(\tau,L)$ decays slowly with a power law towards a constant,
i.e.\ we find $\kappa\propto L^{-9/4}$ independent of $\tau$ for large $\tau$.

\section{Videos of the domain wall motion}
\label{sec:videos}
We provide three videos to exemplify the stochastic domain wall motion and the coupled motion of the ghost particle cloud 
for the cases considered in Fig.~2 of the main text. The density profiles shown in these videos were obtained by 
averaging occupation numbers in time intervals $\Delta t=500$.

\begin{list}{}{\setlength{\leftmargin}{0em}\setlength{\rightmargin}{0em}
\setlength{\itemsep}{0ex}\setlength{\topsep}{0ex}}

\item \textbf{VideoS1}\textit{(.gif)}: Motion of the domain wall and ghost particle cloud for 
$L=500$. The domain wall separates two coexisting maximum current phases with bulk densities $\rhoonemax\cong0.304$ (right side of the domain wall) and $\rhotwomax=1\!-\!\rhoonemax\cong0.696$ (left side of the domain wall). 
The reservoir densities $(\rhoL,\rhoR)=(0.8,0.3016)$ in the coexisting region II+VII of the two extremal current phases 
were chosen to be very close to the boundary to phase VII [see phase diagram in Fig.~1(c) of the main text]. 
As a consequence, the fluctuating domain wall position has a mean value $\langle x\rangle\simeq 102$
shifted to the right of $x=0$, because the phase
with density $\rhoonemax$ (right side) tends to become expelled from the system.

\item \textbf{VideoS2}\textit{(.gif)}: Motion of the domain wall and ghost particle cloud for 
$L=500$ as in VideoS1, but now for reservoir densities $(\rhoL,\rhoR)=(0.8,0.2)$ on the ``symmetry line'' $\rhoR=1-\rhoL$.
The fluctuations of the domain wall position are around the mean position $\langle x\rangle=0$.

\item \textbf{VideoS3}\textit{(.gif)}: Motion of the domain wall and ghost particle cloud for the same reservoir densities $(\rhoL,\rhoR)=(0.8,0.2)$ as in VideoS2, but now for an eight times larger system size $L=4000$.

\end{list}


\end{document}